# Using the Medipix3 detector for direct electron imaging in the range 60keV to 200keV in electron microscopy

J.A. Mir[a,*], R. Plackett[b], I. Shipsey[b] and J.M.F. dos Santos[c]

*a University of Oxford, Department of Materials, Parks Road, Oxford OX1 3PH*
*b University of Oxford, Department of Physics, Parks Road, Oxford OX1 3PH*
*c LIBPhys, Department of Physics, University of Coimbra, P-3004-516 Coimbra, Portugal*

\* Corresponding author. E-mail: jamil.mir@physics.ox.ac.uk

ABSTRACT:

Hybrid pixel sensor technology such as the Medipix3 represents a unique tool for electron imaging. We have investigated its performance as a direct imaging detector using a Transmission Electron Microscope (TEM) which incorporated a Medipix3 detector with a 300 µm thick silicon layer compromising of 256x256 pixels at 55 µm pixel pitch. We present results taken with the Medipix3 in Single Pixel Mode (SPM) with electron beam energies in the range, 60–200 keV. Measurements of the Modulation Transfer Function (MTF) and the Detective Quantum Efficiency (DQE) were investigated. At a given beam energy, the MTF data was acquired by deploying the established knife edge technique. Similarly, the experimental data required to DQE was obtained by acquiring a stack of images of a focused beam and of free space (flatfield) to determine the Noise Power Spectrum (NPS).

# Contents

## 1. Introduction

Direct electron detection can be achieved by using solid state detection technology such as the Monolithic Active Pixel Sensor (MAPS) [1] or a hybrid technology such as the Medipix3 detector [2,3]. The MAPS detectors are well suited for high electron energy regime within the context of imaging performance, although at a substantial reduction in detection efficiency, as a high fraction of the electron beam is transmitted without being detected. Furthermore, at lower electron energies, the imaging performance using MAPS will be reduced due to increased electron backscattering [1]. Beam energies lower than 100 keV can provide greater contrast for thin biological samples [4] or the elimination of knock-on damage by the primary beam, for example in imaging 2-dimensional materials containing light elements such as graphene [5]. For these types of applications, the different architecture of hybrid pixel detectors such as Medipix3 may offer advantages.

The Medipix3 detector was designed at CERN for photon and particle detection using the commercial IBM 0.13 μm CMOS technology and measures 17.3x14.1mm$^2$. The active part of the detector comprises a 300 μm Silicon layer with 256x256 55μm by 55μm square pixels. Each pixel contains analogue and digital circuitry consisting of a charge sensitive preamplifier, a semi Gaussian shaper and two discriminators that control the lower and upper threshold levels. Each discriminator has a 5 bit Digital to Analogue (DAC) to reduce the threshold dispersion caused by mismatches produced during the transistors fabrication. In the standard operating mode, so called Single Pixel Mode (SPM), pixels register a count if it exceeds the preset lower threshold energy value. Each pixel contains two configurable depth registers which function as counters; these have sufficient configurability to allow operation with multiple thresholds, double depth counters or enables a continuous Read-Write capability whereby one register acts as a counter whilst the other shifts the data out.

## 2. Method

We have used a single Medipix3 ASIC/sensor pair to investigate its performance in the 60-200 keV electron energy range. The basic metrics for quantifying the performance of a pixelated detector are the Modulation Transfer Function (MTF) and the Detective Quantum Efficiency (DQE). In the present studies, we have measured the MTF using the established knife edge method [6,7] and DQE by calculating the Noise Power Spectrum (NPS) from stacks of flatfield images. The Medipix3 detector was mounted inside the JEOLARM200cF TEM/STEM. The operation and high-speed data readout of the detector used the MERLIN hardware/software designed by Diamond Light Source and produced by Quantum Detectors [8]. The MTF and DQE data was taken for primary electron beam energies of 60, 80,120 and 200 keV using SPM mode. For each of the electron energies, the MTF data was taken by recording images of a 2 mm thick aluminium knife edge inclined by 10 degrees with respect to the pixel readout columns. Having set the exposure time to 10 ms, 32 repeated images were acquired across the full range of Medipix3 discriminator threshold values, TH0, in the SPM mode.

DQE was measured for a 60, 80, 120 and 200 keV by recording flatfield images to derive the Noise Power Spectrum (NPS), and ultimately the DQE using equation 1. This was accomplished by acquiring a set of 32 flatfield images with 10 ms exposure times across the full range of energy threshold values in the SPM mode. In order to calibrate the threshold discriminator at a given electron energy, the total number of counts in an image were plotted against the discriminator threshold values to form an integral pulse height spectrum. Curve fitting this spectrum and then taking its derivative yielded the normal pulse height distribution of the incident electron energy. The gain factors g for the specified primary electron energies were measured by focusing the electron beam on the detector and recording 32 images at 10 ms as a function of threshold energy values. For each electron energy used, the beam current was measured using a Faraday cup located at the small screen of the JEOLARM200cF TEM/STEM.

3.  Results and Discussion

In order to obtain the MTF for a given electron energy, a Line Spread Function (LSF) was initially obtained by differentiating the experimentally obtained edge profile. The modulus of the Fourier transform of the LSF yields the MTF. As previously reported the DQE can be conveniently calculated from the MTF and the Noise Power Spectrum (NPS) as [7,9]:

$$DQE(f) = \frac{c^2 MTF^2}{n(NPS)} \quad (1)$$

where f is the spatial frequency, c represents the number of counts in the output image, n is the input electron dose and NPS describes the spatial frequency dependence of the noise [7]. Hence, in order to calculate the DQE, experimental measurements of the MTF, NPS and the gain factor, g, defined as the ratio *c/n* for a given electron energy and discriminator threshold are required. The NPS was calculated from the Fourier transform of flatfield images recorded under uniform illumination. MTF and the DQE were evaluated in the spatial frequency range from 0 to 0.5 pixel$^{-1}$ where the upper limit represents the Nyquist frequency beyond which aliasing occurs.

Figure 1 shows the variation of the MTF as a function of spatial frequency at 60, 80, 120 and 200 keV electrons using the SPM mode. The discriminator threshold was set just above noise at 5.3 keV. It is evident that there is a progressive degradation in the MTF with increasing electron energy. Figure 2 shows the log-normal relationship between the MTF and DQE at the Nyquist frequency and electron energy. The reduction in the MTF is attributed to the increased electron scattering range leading to distant pixels being triggered with respect to the initial impact point. This was verified by Monte Carlo simulations carried out using the CASINO software package [10] with silicon substrate thickness set at 300 μm. These simulations showed that the lateral charge spread (95%) at 60 keV is approximately 24μm and increases to approximately 39 μm, 83μm and 170 μm at 80, 120 and 200 keV, respectively.

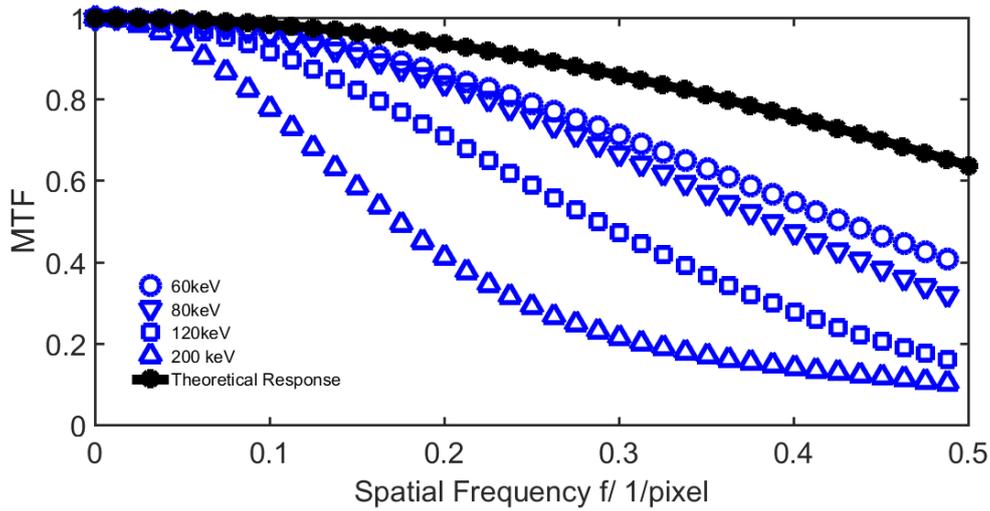

Figure 1: MTF as a function of the spatial frequency at 60, 80, 120 and 200 keV for Single Pixel Mode (SPM) with the discriminator threshold set just above noise at 5.3 keV. The theoretical response of a detector with square pixels is also shown.

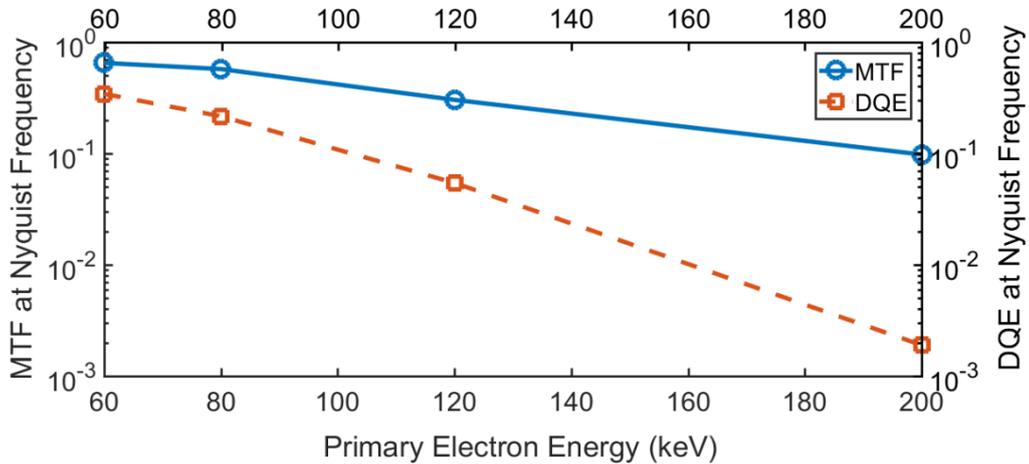

Figure 2: Variation of the MTF and DQE at the Nyquist frequency as a function of electron energy using the Single Pixel Mode (SPM). The discriminator threshold at a given energy was set at half the primary electron energy.

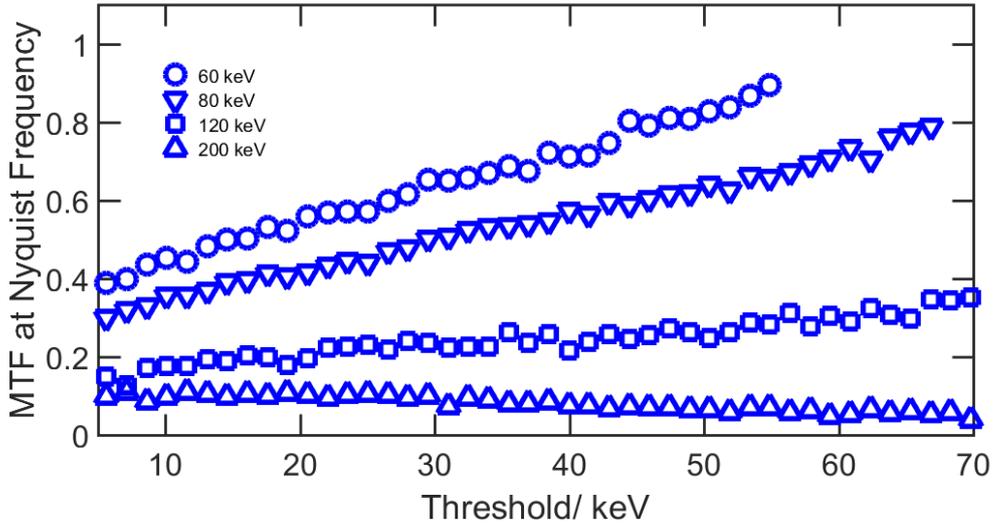

Figure 3: Comparison of MTF at the Nyquist frequency as a function of discriminator threshold using the Single Pixel Mode (SPM) at 60, 80, 120 and 200 keV.

Figure 3 shows the variation of MTF at the Nyquist frequency as a function of the discriminator threshold at 60, 80, 120 and 200 keV using SPM. The MTF for a theoretical square pixel detector at Nyquist frequency equals 0.64 (see Figure 1) which is experimentally achieved for the 60 keV alone since the maximum lateral dispersion at this energy does not exceed half the pixel width when the when the discriminator threshold is set at 30 keV. For the discriminator threshold set below 30 keV, multiple pixels are triggered leading to MTF degradation. Conversely, for discriminator threshold set above 30 keV, MTF is better than the theoretical maximum due to the reduction in the effective pixel size [11].

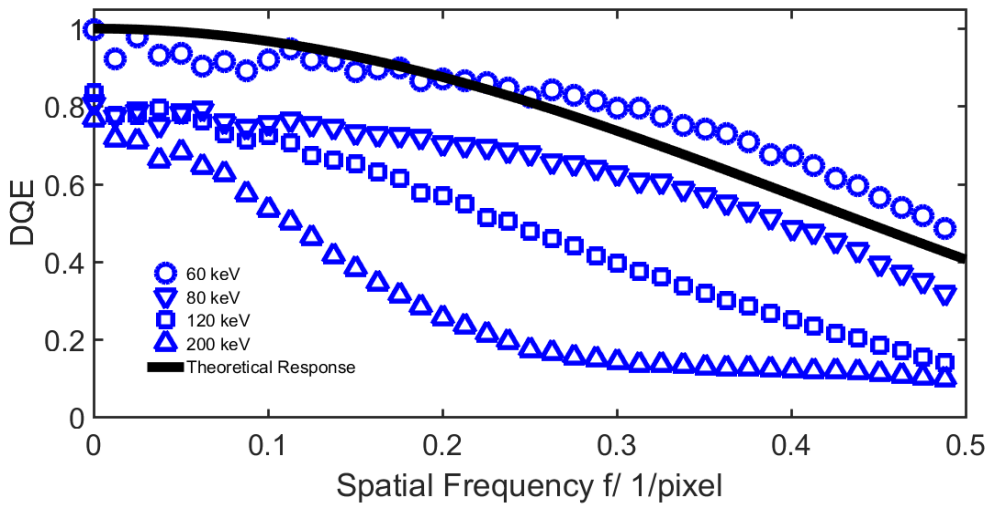

Figure 4: DQE as a function of the spatial frequency at 60, 80, 120 and 200 keV for Single Pixel Mode (SPM) with discriminator threshold set just above noise at 5.3 keV. The theoretical (ideal) response of a detector with square pixels is also shown as a solid line.

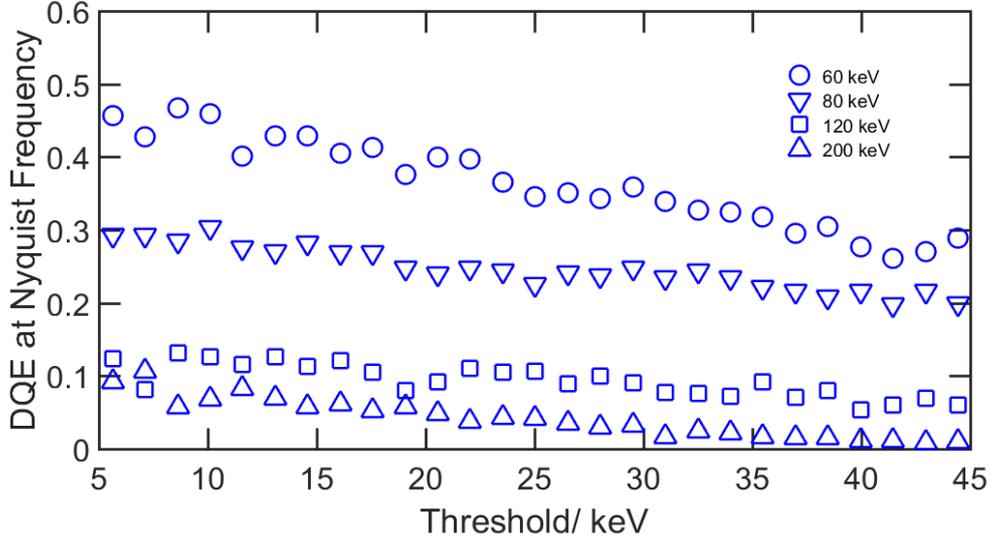

Figure 5: DQE at the Nyquist frequency as a function of discriminator threshold using the Single Pixel Mode (SPM) at 60, 80, 120 and 200 keV.

The reduction in MTF with increasing electron energy impacts DQE proportionally (see equation 1) as shown in Figures 4 and 5. As the electron scattering cross sections decrease with increasing electron energy, a smaller proportion of electrons are detected. The DQE for a theoretical square pixel detector at the Nyquist frequency equals 0.41 (see Figure 4) which is experimentally achieved for the 60 keV electrons only with progressive degradation at higher energies.

## 4. Conclusions

We have investigated the imaging response of the Medipix3 sensor at 60, 80, 120 and 200 keV electron energies. In SPM mode, the MTF exhibits a direct dependence on energy threshold where high discriminator thresholds yield higher MTFs. Conversely, the DQE exhibits a strong inverse dependence on the energy threshold where low discriminator thresholds yield the higher DQEs. These measurements agree with trends already observed for the Medipix2 detector with the exception of performance at 200 keV where the MTF and DQE are insensitive to the variations in the discriminator threshold [11].

Primary electron energies of 60-80 keV range are highly relevant in the imaging of novel 2D materials such as graphene [5] for which we have shown that operating the Medipix3 detector in the SPM mode is sufficient. At lower electron energies, the MTF and DQE were found to approach values corresponding to those of a theoretical square pixel detector when the energy threshold was set at half the primary electron energy. For example, the MTF and the DQE at 60 keV at the Nyquist frequency were found to be 0.64 and 0.36, respectively with the energy threshold set at 30 keV. For comparison, the MTF and DQE for a theoretical square pixel detector at the Nyquist frequency equal 0.64 and 0.41, respectively.

## Acknowledgments

We would like to thank Prof. Val O' Shea of Glasgow University and his group for access to the JEOLARM200cF TEM/STEM. Financial support from the European Union under the Seventh Framework Program under a contract for an Integrated Infrastructure Initiative (Ref 312483-ESTEEM2) is gratefully acknowledged.